\documentclass[twocolumn,showpacs,secnumarabic,superscriptaddress,nobibnotes,aps,pre]{revtex4-2}
\usepackage{amssymb,amsmath,graphicx,listings}
\usepackage{mathrsfs}
\usepackage{url,hyperref}
\usepackage[usenames,dvipsnames]{xcolor}
\hypersetup{colorlinks=true, linkcolor=BrickRed, urlcolor=blue!50!black, citecolor=blue!50!black}
\usepackage[capitalize]{cleveref}
\usepackage{cleveref}
\usepackage{multirow}
\usepackage{float}
\AtBeginDocument{}
\crefrangelabelformat{equation}{(#3#1#4)$-$(#5#2#6)}
\newcommand{\eqn}[1]{\begin{equation} #1 \end{equation}}

\begin{document}
\title{Critical Fluctuations at Finite-Time Dynamical Phase Transition}
\author{Nalina Vadakkayil}
\affiliation{Complex Systems and Statistical Mechanics, Department of Physics and Materials Science, University of Luxembourg, L-1511 Luxembourg, Luxembourg}
\author{Massimiliano Esposito}
\affiliation{Complex Systems and Statistical Mechanics, Department of Physics and Materials Science, University of Luxembourg, L-1511 Luxembourg, Luxembourg}
\author{Jan Meibohm}
\affiliation{Technische Universit\"at Berlin, Stra\ss{}e des 17. Juni 135, 10623 Berlin, Germany}
\affiliation{Department of Mathematics, King's College London, London WC2R 2LS, United Kingdom}
\date{\today}

\begin{abstract}
We explore the critical properties of the recently discovered finite-time dynamical phase transition in the non-equilibrium relaxation of Ising magnets after a temperature quench. The transition is characterized by a sudden switch in the relaxation dynamics and it occurs at a sharp critical time. While previous works have focused either on mean-field interactions or on investigating the critical time, we analyze the critical fluctuations at the phase transition in the nearest-neighbor Ising model on a square lattice using Monte Carlo simulations.  By means of a finite-size scaling analysis, we extract the critical exponents for the transition. In two spatial dimensions, we find that the exponents are consistent with those of the two-dimensional Ising universality class when the system is initially in the vicinity of the critical point. For initial temperatures below the critical one, however, the critical exponents differ from the Ising-exponents, indicating a distinct dynamical critical phenomenon.
\end{abstract}
\maketitle
\section{Introduction}\label{sec:intro}
Non-equilibrium phase transitions are common across a wide range of disciplines in science~\cite{haken2004introduction,henkel2011non}, ranging from biological systems to galactic patterns. Their non-equilibrium nature allows these transitions to  manifest themselves in a rich variety of both static and dynamic patterns~\cite{cross1993pattern}. Equilibrium phase transitions, by contrast, are more limited in regard to the possible ways in which they may occur. Furthermore, universality, a common property of equilibrium phase transitions, constrains the critical properties of phase transitions at equilibrium. As a consequence of these restrictions, equilibrium phase transitions are theoretically tractable and are nowadays comprehensively understood within the theory of equilibrium statistical mechanics~\cite{landau2012course,goldenfeld}. Similarly, near-equilibrium systems are in the realm of powerful theoretical frameworks, such as linear-response theory, that build upon the equilibrium theory.

Far from equilibrium, no theory of comparable scope is known and although some equilibrium methods can be generalized to far-from-equilibrium~\cite{marconi,baiesi2013}, the lack of an overarching theory makes the study of non-equilibrium phase transitions challenging.

Far-from-equilibrium thermal relaxation is a prime example of a genuine non-equilibrium process. As such, it exhibits a number of anomalous features, including ergodicity breaking~\cite{bray,henkel2011non}, the Mpemba~\cite{mpemba,lu2017nonequilibrium,vadakkayil2021} and Kovacs effects~\cite{kovacs1963dynamic,kovacs,bertin2003kovacs}, and asymmetries associated with heating and cooling processes~\cite{lapolla2020,meibohm2021,ibanez2024heating}. Such anomalous relaxation phenomena are often discussed in connection with an associated equilibrium phase transition~\cite{holtzman2022}.

Recent studies~\cite{jan_massi_prl, jan_massi_njp, BlomGodec2023} revealed anomalous behavior in the extreme-event statistics of non-equilibrium relaxation in Ising magnets, namely, so-called finite-time dynamical phase transitions. These transitions manifest themselves through the emergence of finite-time kinks in the large-deviation (rate) functions of thermodynamic observables within the Curie-Weiss model of $N$ globally coupled Ising spins~\cite{jan_massi_prl, jan_massi_njp} -- see also the mathematical works in Refs.~\cite{van2002possible,kulske2007spin,ermolaev2010low,redig2012gibbs}. The transitions occur at sharp, ``critical'' times during the transient relaxation following an instantaneous quench from the ferromagnetic phase into the paramagnetic phase of the model. Remarkably, many key features of this non-equilibrium transition are identical to the equilibrium phase transition of the Curie-Weiss model, including its (mean-field) critical exponents~\cite{jan_massi_prl, jan_massi_njp}. Some effort has been made to explore this transition to models with short-range interactions~\cite{BlomGodec2023}. So far, however, these studies have focused on accurately determining the critical time.

Here, we investigate critical phenomena associated with the finite-time dynamical phase transition in an Ising model with nearest-neighbor interactions on a two-dimensional square lattice of linear dimension $L$ with periodic boundary conditions. Although we do not expect equilibrium-like universality, categorizing critical fluctuations close to non-equilibrium phase transitions in terms of their critical exponents helps to characterize these transitions~\cite{tauber2014,baglietto2008finite,wood2006universality,wood2006critical}. In some systems that exhibit both equilibrium and non-equilibrium phase transitions, the critical properties of the non-equilibrium transitions have been shown to fall in the same universality class as the corresponding equilibrium transitions~\cite{wood2006universality}, leading to interesting correspondences between equilibrium and non-equilibrium.

Using a finite-size scaling analysis \cite{fisher_barber}, we calculate the critical exponents associated with the magnetization and the magnetic susceptibility. We show that the values of these exponents are consistent with those of the two-dimensional Ising universality class at equilibrium, when the initial temperature approaches the critical point. This behavior is akin to that of the mean-field model, where the critical exponents of finite-time dynamical phase transitions are identical to those of their equilibrium counter parts. For initial temperatures away from the critical point, however, we find that the critical exponents depart from the equilibrium values. This indicates that, on the lattice, finite-time dynamical phase transitions are non-equilibrium transitions in their own right, whose critical properties are generally distinct from those at equilibrium.

The paper is organized as follows: In Sec.~\ref{sec:background}, we discuss the model and our methodology, and explain them by reviewing the well-known equilibrium phase transition in the two-dimensional Ising model. We also review the occurrence of the finite-time dynamical phase transitions in the mean-field version of the model. In Sec.~\ref{sec:results}, we present our main results characterizing the critical properties of the nearest-neighbor Ising model. We draw our conclusions in Sec.~\ref{sec:conclusion}. The numerical values for the critical time and the critical exponents are summarized in Appendix A.
\section{Background}\label{sec:background}
In this Section, we introduce the model and discuss its equilibrium properties and how to treat finite-size effects. Subsequently, we review the origin of the finite-time dynamical phase transition in the mean-field version of the model~\cite{jan_massi_prl, jan_massi_njp}.
\subsection{Model}\label{sec:model}
We explore the critical properties of the finite-time dynamical phase transition for the nearest-neighbor Ising model (\textbf{NNIM}) on a two-dimensional square lattice with periodic boundary conditions. The internal energy of the model at vanishing external magnetic field reads \cite{bray,fisher_scaling}
\eqn{\label{eq:Hamiltonian}
 E = -J \sum_{\langle ij\rangle} S_iS_j\,,
}
where $J>0$ is the ferromagnetic nearest-neighbor coupling and $S_{i}=\pm 1$ denotes the state of the Ising spin at lattice site $i=1,\ldots,N$. The sum in Eq.~\eqref{eq:Hamiltonian} runs over all pairs of nearest neighbors.

In the mean-field version of the model, briefly discussed in Sec.~\ref{sec:ftdpt}, one sums over all pairs of lattice sites (instead of nearest neighbors), and the coupling $J$ is scaled with the number $N$ of spins \cite{griffiths1966relaxation}.

The system is immersed in a heat bath at inverse temperature $ 1/(k_B T)$. At equilibrium, the NNIM exhibits a second-order phase transition from a disordered into an ordered phase as a function of the so-called inverse coupling temperature,
\eqn{
	\mathscr{J}:= J/(k_BT)\,.
}
The equilibrium transition occurs at the critical value $\mathscr{J}_c := J/(k_BT_c)$, where $T_c$ denotes the Curie temperature. For the two-dimensional NNIM, the value of $\mathscr{J}_c$ is given by~\cite{onsager1944}
\eqn{\label{eq:Jc}
	\mathscr{J}_c  = \frac{\ln(1+\sqrt 2)}{2}\approx0.44\,,
}
while it is equal to unity in the mean-field version of the model. 
\subsection{Dynamics}\label{sec:dynamics}
The interaction of the system with the heat bath generates stochastic spin-flips that obey detailed balance. Randomly selected spins on the lattice are flipped according to the Glauber dynamics~\cite{glauber} where the transition rate for a spin flip on a site is given by
\eqn{\label{eq:glauber}
W_\text{flip} = \frac{\tau^{-1}}{e^{\frac{\Delta E}{k_\text{B}T}}+1}\,,
}
where $\Delta E$ denotes the energy change resulting from the flip, $k_\text{B}$ is Boltzmann's constant, $T$ denotes the temperature of the surrounding heat bath, and $\tau$ is the characteristic time scale of a single spin-flip. Since the transition rate in Eq.~\eqref{eq:glauber} obeys detailed balance, the system eventually reaches equilibrium at inverse coupling temperature $\mathscr{J}$ in the long-time limit. Using Glauber dynamics with the rate Eq.~\eqref{eq:glauber}, we evolve the state of the NNIM by discrete-time Monte Carlo simulations~\cite{landau_binder,newman} with periodic boundary conditions. In our simulations, $N$ attempts to flip a spin constitute a single Monte Carlo step, which in turn corresponds to an advance by $\tau$ in physical time.
\subsection{Equilibrium phase transition}
\begin{figure*}[htp]
  \includegraphics*[width=\linewidth]{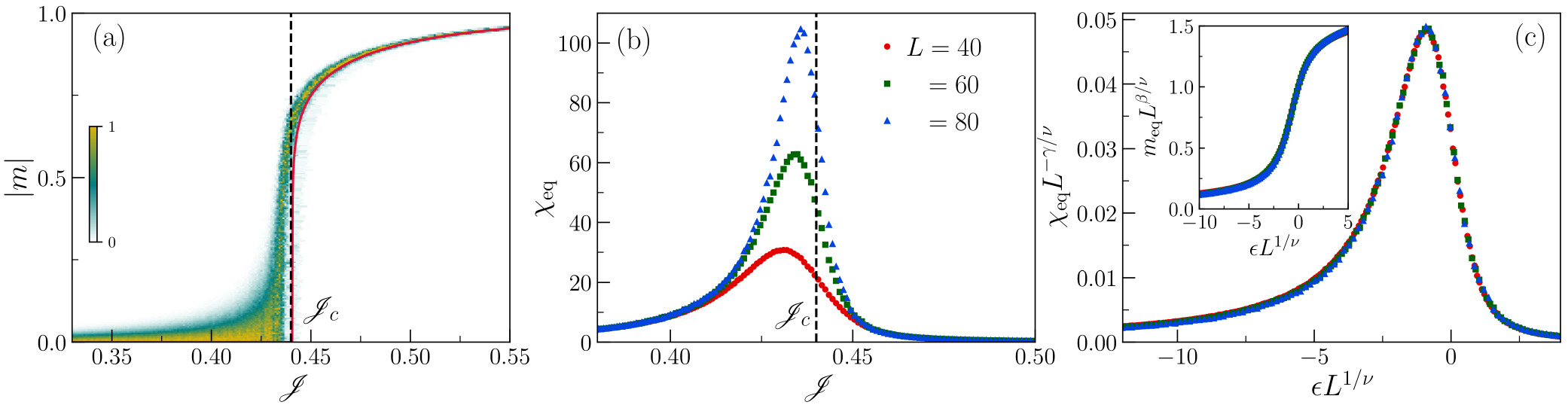}
  \caption{Equilibrium (para- to ferromagnetic) phase transition of the model, obtained from Monte Carlo simulations with a minimum of $10^6$ independent initial configurations (see text). (a) Probability density of $|m|$ (heat map) as a function of $\mathscr{J}$ for $L=200$. The solid line shows the exact result in Eq.~\eqref{eq:meqons} for $m_\text{eq}$, valid as $L\to\infty$. The dashed line marks the critical point, $\mathscr{J}_c$ given in Eq.~\eqref{eq:Jc}. (b) Magnetic susceptibility, $\chi_\text{eq}$, as a function of $\mathscr{J}$ for different system sizes. (c) Collapse of $\chi_\text{eq}$ (main panel) and $m_\text{eq}$ (inset) onto the scaling functions $Y_{\chi_\text{eq}}$ and $Y_{m_\text{eq}}$, respectively (see text).}
  \label{fig:equilibrium}
\end{figure*}
Before discussing the finite-time dynamical phase transition in more detail, we review the equilibrium phase transition of the NNIM. For the Monte Carlo simulations of the model  at equilibrium, we employ the Wolff algorithm~\cite{wolff1989collective}, a non-dynamical algorithm that ensures fast convergence to the equilibrium state, even in the vicinity of the critical point.

An equilibrium phase transition is characterized by a qualitative change of the equilibrium state of a system in response to a quasistatic change of the external parameters~\cite{landau2012course,goldenfeld}. In particular, when $\mathscr{J}$ is increased above $\mathscr{J}_c$, the NNIM transitions from a disordered paramagnetic phase to an ordered ferromagnetic phase. The order parameter for this transition is the magnetization $m$ per spin, defined as
\eqn{\label{eq:magnetization}
    m := L^{-d}\sum_{i=1}^NS_i\,,
}
where $d$ denotes the lattice dimension with $d=2$ here and in the following, unless stated otherwise. The spontaneous magnetization $m_\text{eq}(L)$ at finite system size $L$ is obtained from
\eqn{\label{eq:meq}
	m_\text{eq}(L) := \langle |m| \rangle\,,
}
where the brackets $\langle\ldots\rangle$ denote an average over an ensemble of equilibrium systems. Taking the thermodynamic limit in two spatial dimensions, $m_\text{eq}(L)$ converges to the exact expression computed by Onsager~\cite{onsager1944}:
\eqn{\label{eq:meqons}
	m_\text{eq}:=\lim_{L\to\infty}m_\text{eq}(L) = [1 - \sinh^{-4}(2\mathscr{J})]^{1/8}\,.
}

As its role as an order parameter suggests, $m_\text{eq}$ vanishes in the paramagnetic phase ($\mathscr{J}<\mathscr{J}_c$), while $m_\text{eq}>0$ in the ferromagnetic phase ($\mathscr{J}>\mathscr{J}_c$).

The solid line in Fig.~\ref{fig:equilibrium}(a) shows the spontaneous magnetization $m_\text{eq}$ [Eq.~\eqref{eq:meqons}] as a function of $\mathscr{J}$ for the two-dimensional NNIM. The heat map represents the probability density of $|m|$ at finite system size, normalized to unity for each $\mathscr{J}$. We observe that at fixed $\mathscr{J}$ and for large $L$, the spontaneous magnetization $m_\text{eq}$ is well approximated by the regions of the largest probability density.

The equilibrium phase transition in the NNIM is continuous, meaning that $m_\text{eq}$ changes continuously at the critical point. In the vicinity of this point, several thermodynamic quantities diverge in the thermodynamic limit and exhibit the so-called scaling behavior~\cite{fisher_scaling}. For example, close to the critical point, we obtain from Eq.~\eqref{eq:meqons} that $m_\text{eq}$ behaves as
\eqn{\label{eq:beta_eq}
m_\text{eq} \sim \epsilon^{\beta}\,,\qquad \epsilon := \frac{\mathscr{J} - \mathscr{J}_c}{\mathscr{J}}\,,
}
for $0<\epsilon\ll1$, with critical exponent $\beta = \frac18$~\cite{fisher_scaling,stanley}.

Furthermore, the magnetic susceptibility $\chi_\text{eq}$ [in units of $1/(k_B T)$] is a measure the magnitude of fluctuations of the magnetization around its mean, i.e.,
\eqn{\label{eq:susc}
	\chi_\text{eq} := L^{d}\left(\langle m^2\rangle - \langle |m|\rangle^2\right)\,.
}
Close to the critical point, $\chi_\text{eq}$ diverges as
\eqn{\label{eq:chi_eq}
	\chi_\text{eq} \sim |\epsilon|^{-\gamma}\,,
}
where $\gamma = \frac74$ is the associated critical exponent~\cite{fisher_scaling,stanley}.

The exponents $\beta$ and $\gamma$ have universal values across a range of systems, based on their microscopic symmetries, giving rise to a so-called universality class. Within the two-dimensional Ising universality class, relevant here, one has $\beta=\frac{1}{8}$ and $\gamma=\frac{7}{4}$ as stated above, while within the mean-field universality class, the corresponding values are $\beta = \frac12$ and $\gamma = 1$ \cite{fisher_scaling,stanley}.
\subsection{Finite-size scaling}\label{sec:fss}
The power-law scalings in Eqs.~\eqref{eq:beta_eq} and \eqref{eq:chi_eq} are properties of the infinite system. By contrast, whenever the NNIM is studied by means of simulations, one needs to resort to finite systems. Operating at finite system size changes the behavior close to the critical point due to so-called finite-size effects \cite{privman1990}.

Figure~\ref{fig:equilibrium}(b) shows a prime example of a finite-size effect that occurs in the magnetic susceptibility $\chi_\text{eq}$ of the NNIM at equilibrium. As a function of the inverse coupling temperature $\mathscr{J}$, $\chi_\text{eq}$ does \textit{not} diverge, as it would according to the predicted behavior in Eq.~\eqref{eq:chi_eq}, but exhibits a characteristic peak that becomes more and more pronounced as the system size increases. The location of the peak serves as a finite-size estimate of the critical inverse coupling temperature $\mathscr{J}_c(L)$~\cite{privman1990}.

Although the divergence of $\chi_{\rm eq}$ in Eq.~\eqref{eq:chi_eq} and the limit $\lim_{L\to\infty}\mathscr{J}_c(L) = \mathscr{J}_c$ are not directly observable in any finite system, analyzing the finite-size scaling~\cite{fisher_scaling,fisher_barber,landau_binder,newman} of the system allows one to infer the limiting properties from size-dependent measurements.

Fortunately, for large-enough systems at the critical point, finite-size scaling simplifies, because the correlation length, $\xi_\text{eq}$, over which the spins are correlated~\footnote{The correlation length $\xi_\text{eq}$ is the characteristic decay length of the two-point correlation function, $C(i,j) := \langle S_i S_j \rangle -\langle S_i\rangle \langle S_j\rangle$~\cite{bray}. Near the critical point, $C(i,j)$ decays as $\propto e^{-|i-j|/\xi_\text{eq}}$.}, is the only relevant length scale~\cite{privman1990}.  In the thermodynamic limit, $\xi_\text{eq}$ diverges at the critical point as,
\eqn{\label{corrlenghtdiv}
    \xi_\text{eq} \sim |\epsilon|^{-\nu}\,,
}
with an associated exponent $\nu$, whose value is $\nu=1$ within the two-dimensional Ising universality class and $\nu = \frac12$ for mean-field models \cite{fisher_scaling}.
 
In finite systems, however, the correlation length $\xi_\text{eq}$ is bounded, i.e., cut off, by the system size $L$ at the size-dependent critical inverse coupling temperature $\mathscr{J}_c(L)$. Using \eqref{corrlenghtdiv}, we can therefore rewrite~\eqref{eq:beta_eq} and \eqref{eq:chi_eq} in terms of $L$ close to the critical point,
\eqn{\label{eq:mag_eq_scaling}
	m_\text{eq}(L) \sim L^{-\beta/\nu},\qquad \chi_\text{eq}(L) \sim L^{\gamma/\nu}\,,
}
for $0<\epsilon\ll1$.

Finite-size effects in $m_\text{eq}(L)$ and $\chi_\text{eq}(L)$ can be summarized by introducing two corresponding scaling functions that are constructed to remain finite at the critical point for any $L$. The first one reads
\eqn{\label{eq:beta_eq_scaling}
    Y_{m_\text{eq}}(y) := m_\text{eq}(L) L^{\beta/\nu}\,,
}
where $y$ is a dimensionless scaling variable given by 
\eqn{\label{eq:scalvar}
	y := \epsilon L^{1/\nu}\sim\text{sign}(\epsilon)\left(\frac{L}{\xi_\text{eq}}\right)^{1/\nu}\,,
}
and the second one is obtained as 
\eqn{\label{eq:chi_eq_scaling}
Y_{\chi_\text{eq}}(y) := \chi_\text{eq}(L) L^{-\gamma/\nu}\,.
} 

The use of scaling functions is that, provided the correct values of the critical exponents $\beta$, $\gamma$, and $\nu$ are known, the rescaled finite-size curves of $m_\text{eq}(L) L^{\beta/\nu}$ and $\chi_\text{eq}(L) L^{-\gamma/\nu}$ collapse onto $Y_{m_\text{eq}}(y)$ and $Y_{\chi_\text{eq}}(y)$, respectively.

Figure~\ref{fig:equilibrium}(c) shows the collapse of the finite-size equilibrium data onto $Y_{\chi_\text{eq}}$ and $Y_{m_\text{eq}}$ (inset), using the known values for the critical exponents of the two-dimensional Ising universality class stated above. The excellent collapse of the data in Fig.~\ref{fig:equilibrium}(c) confirms that the scaling assumptions together with the known values of critical exponents are consistent with the equilibrium data.
\begin{figure*}[htp]
\centering
  \includegraphics*[width=\linewidth]{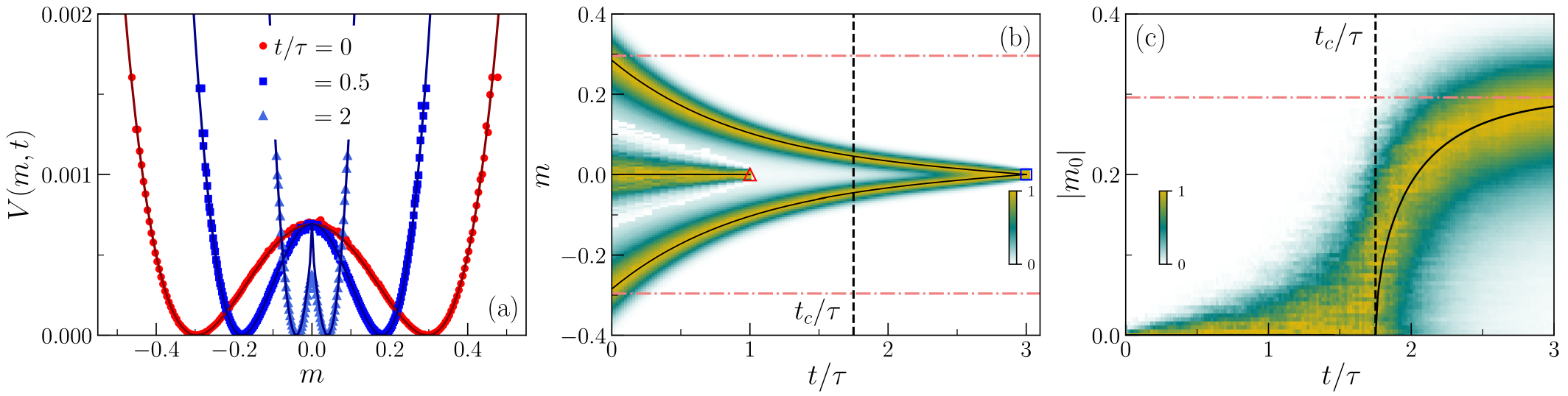}
  \caption{Finite-time dynamical phase transition of the mean-field model for $L = 80$, obtained from $10^6$ realizations of the relaxation process. (a) Time-dependent rate function $V(m,t)$ as a function of $m$ after a quench from $\mathscr{J} = 1.03\mathscr{J}_c$ to $\mathscr{J}_q = 0$. The red line (theory) and red bullets (simulations) show the equilibrium rate function $V_\text{eq}$ for $\mathscr{J} = 1.03\mathscr{J}_c$. The blue lines (theory) and blue symbols (simulations) show $V(m,t)$ at different conditioning times, $t/\tau = 0.5$ and $2$. (b) Optimal fluctuations before and after the transition, obtained from $5\times10^3$ and $7\times10^5$ conditioned relaxation trajectories, respectively. The red triangle and the blue square show the endpoints of the conditioned trajectories before and after the critical time (dashed line). The solid lines are the optimal fluctuations obtained from theory. The horizontal dash-dotted lines show the initial spontaneous magnetizations $\pm m_\text{eq}$ for $\mathscr{J} = 1.03 \mathscr{J}_c$. (c) Probability density of $|m_0|$ at different times, obtained from between $\sim 10^2$ (smallest $t/\tau$) and $\sim10^5$ (largest $t/\tau$) conditioned relaxation trajectories. The solid lines are obtained from mean-field calculations~\cite{jan_massi_prl}. The dashed line shows the critical time $t_c$ and the dash-dotted line shows the initial spontaneous magnetizations $\pm m_\text{eq}$.}
  \label{fig:rates_mf}
\end{figure*}
\subsection{Large-deviation theory}
Finite-size fluctuations, both at the critical point and away from it, can be reformulated using large-deviation theory~\cite{touchette2009large}. At large but finite system size, the magnetization $m$ exhibits equilibrium fluctuations. Away from the critical point, the probability density of these fluctuations attains a so-called large-deviation form~\cite{touchette2009large} as $L\to\infty$:
\eqn{\label{eq:rfeq}
	P_\text{eq}(m) \asymp e^{-L^d V_\text{eq}(m)}\,,
}
where $V_\text{eq}(m)$ denotes the equilibrium rate function, a central object in the theory of large deviations. Here, $V_\text{eq}(m)$ measures the exponential rate at which the probability that $m$ takes a given value tends to zero as $L\to\infty$. At the typical (i.e. most likely) event $m=m_\text{eq}$, $V_\text{eq}(m)$ vanishes, $V_\text{eq}(m_\text{eq})=0$. The magnetic susceptibility $\chi_\text{eq}$, in turn, is given by the inverse of the curvature of $V_\text{eq}(m)$ at $m=\pm m_\text{eq}$, i.e.,
\eqn{
	\chi_\text{eq} = \frac1{V''_\text{eq}(m_\text{eq})}\,.
}
While $m_\text{eq}$ and $\chi_\text{eq}$ characterize the mean value and Gaussian fluctuations of $m$, respectively, $V_\text{eq}(m)$ also characterizes the leading-order scaling of the probability of large deviations far away from $m=\pm m_\text{eq}$.

In the vicinity of the critical point, $\chi_\text{eq}$ diverges as the system size tends to infinity so that the formulation in Eq.~\eqref{eq:rfeq} breaks down. Instead, critical fluctuations of the magnetization are expressed using the rescaled magnetization $\tilde m = m L^{\beta/\nu}$, whose fluctuations are of order unity. As the limit of infinite system size is approached, critical fluctuations are characterized by a universal probability distribution \cite{binder1981critical,balog2022critical}
\eqn{\label{eq:puniv}
	P_\text{eq}(\tilde m,y) \asymp e^{-V_y(\tilde m)}\,,
}
characterized by a scale-invariant rate function $V_y(\tilde m)$ that depends on the scaling variable $y$ defined in Eq.~\eqref{eq:scalvar}, but is otherwise independent of the system size.
\subsection{Finite-time dynamical phase transition}\label{sec:ftdpt}
Having discussed the equilibrium phase transition of the NNIM in some detail, we proceed to the non-equilibrium scenario. We perform an instantaneous disordering quench of the inverse coupling temperature $\mathscr{J}\to\mathscr{J}_q$ from the ordered phase ($\mathscr{J} > \mathscr{J}_c$) into the disordered phase ($\mathscr{J}_q < \mathscr{J}_c$). The quench drives the system far from equilibrium. In the transient that follows, the system relaxes to the new equilibrium state at inverse coupling temperature $\mathscr{J}_q$.

During the transient the system evolves with the dynamics described in Sec.~\ref{sec:dynamics}. We analyze the evolution of the probability density $P(m,t)$ of the magnetization $m$ as a function of time $t$. The large-deviation form~\eqref{eq:rfeq} then becomes time dependent, i.e.,
\eqn{\label{eq:rfneq}
	P(m,t) \asymp e^{-L^d V(m,t)}\,,
}
with time-dependent rate function $V(m,t)$. Similar to the equilibrium rate function, $V_\text{eq}(m)$, in Eq.~\eqref{eq:rfeq}, $V(m,t)$ measures the exponential rate at which the probability that the magnetization takes the value $m$ at a given time $t$ after the quench tends to zero as $L\to\infty$. 

In the mean-field version of the Ising model, the rate function $V(m,t)$ of the magnetization and of other thermodynamic observables form kink-shaped singular points after sharp finite times \cite{jan_massi_prl,jan_massi_njp} in response to the disordering quench. At vanishing magnetic field, the kink in $V(m,t)$ forms at $m=0$ at the critical time~\cite{ermolaev2010low,jan_massi_prl},
\eqn{\label{eq:tc_mf}
	t^\text{MF}_c = \frac{\tau}{2(\mathscr{J}_c - \mathscr{J}_q)}\ln \left(\frac{\mathscr{J}-\mathscr{J}_q}{\mathscr{J}-\mathscr{J}_c}\right)\,.
}

Figure~\ref{fig:rates_mf}(a) shows the evolution of $V(m,t)$ for the mean-field model. The symbols mark the results of Monte Carlo simulations, and the solid lines are exact results from mean-field calculations~\cite{jan_massi_prl}. The initial rate function $V(m,0) = V_\text{eq}(m)$ has a characteristic double-well shape, with two characteristic minima that correspond to the typical values $m=\pm m_\text{eq}$. As time evolves, the minima of $V(m,t)$ approach the origin and a sharp kink forms at $m=0$. The finite values of the time-dependent rate function $V(0,t)>0$ at $m=0$ imply that the kink is associated with large deviations of the magnetization, generated by rare relaxation events, where the system relaxes to vanishing magnetization, $m=0$, in a short time.

The formation of the kink in $V(m,t)$ can be understood as a consequence of competing relaxation dynamics within the system that achieve $m=0$ at a short time $t$: At $t=t^\text{MF}_c$, the most likely way in which $m=0$ is realized, i.e., the so-called optimal fluctuation, suddenly switches, resulting in the finite-time dynamical phase transition. We briefly review this mechanism in the following.

Figure~\ref{fig:rates_mf}(b) shows the optimal fluctuations for $m=0$ in terms of their trajectory densities at different conditioning times $t$ before and after $t^\text{MF}_c$. The normalized densities of conditioned relaxation trajectories are shown as a heat map with yellow regions corresponding to maximum density. Ensembles of conditioned trajectories are obtained by simulating a large number of relaxation trajectories ($\sim10^7$), but keeping only those for which $m$ is within a small window $m=[-\text{d}m,\text{d}m]$ of size $\text{d}m =0.3 N^{-1/2}$ at a given time $t$. After this conditioning, the number of trajectories is substantially reduced to typically between $\sim10^3$ (small $t$) and $\sim10^5$ (larger $t$). The solid lines in Fig.~\ref{fig:rates_mf}(b) are exact results obtained from mean-field computations~\cite{jan_massi_prl}. The dashed vertical line marks $t^\text{MF}_c$, calculated from Eq.~\eqref{eq:tc_mf}. The symbols mark the (fixed) endpoints of the conditioned trajectories at two different times before and after the transition.

We observe a marked difference between the typical relaxation to $m=0$ before and after $t^\text{MF}_c$: At times smaller than the critical time, $t < t^\text{MF}_c$, the trajectories that realize $m=0$ are most likely to initiate at $m=0$ at time $t=0$. Although this initial condition is exponentially unlikely, its occurrence is probabilistically favored compared to trajectories that initiate at a more likely initial magnetization (e.g. close to the spontaneous magnetization) but then require many coordinated spin flips to arrive at $m=0$ at time $t$. For times larger than $t_c$, by contrast, the situation is reversed. The trajectories that achieve $m=0$ at time $t$ are now most likely to initiate from $m\approx \pm m_\text{eq}$, shown as horizontal dash-dotted lines in Fig.~\ref{fig:rates_mf}(b), because now a moderate rate of spin flips suffices to reach $m=0$ in the given time window. 

\begin{figure*}[htp]
\centering
  \includegraphics*[width=\linewidth]{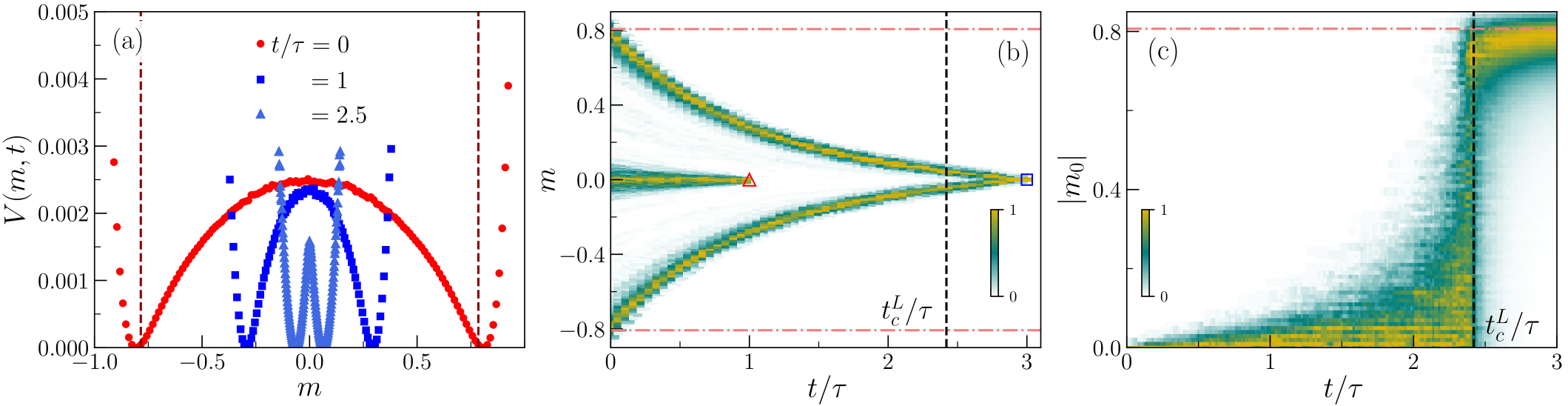}
  \caption{Finite-time dynamical phase transitions in NNIM for $L=60$, obtained from $3\times10^6$ realizations of the relaxation process. (a) Time-dependent rate function $V(m,t)$ as a function of $m$ after a quench from $\mathscr{J} = 1.03\mathscr{J}_c$ to $\mathscr{J}_q = 0$. The curve with red symbols shows the initial rate function $V_\text{eq}(m)$ for $\mathscr{J} = 1.03\mathscr{J}_c$. The curves with blue symbols show $V(m,t)$ at different conditioning times, $t/\tau = 1$ and $2.5$. The dashed lines are the spontaneous magnetization at equilibrium for $\mathscr{J} = 1.03\mathscr{J}_c$. (b) Optimal fluctuations for the NNIM before and after the transition, obtained from $10^3$ and $10^4$ conditioned relaxation trajectories, respectively. The red triangle and the blue square show the endpoints of the trajectories before and after the critical time (dashed line). (c) Probability density of $|m_0|$ at different times, obtained from between $\sim 10^2$ (smallest $t/\tau$) and $\sim 10^5$ (largest $t/\tau$) conditioned trajectories. The dashed line marks the critical time. In both (b) and (c) the horizontal dash-dotted lines represent the minima of the equilibrium rate function, $\pm m_{\rm{eq}}$, corresponding to $\mathscr{J} ( = 1.03\mathscr{J}_c)$.}
  \label{fig:rates_nn}
\end{figure*}

To precisely detect and characterize this transition one analyzes the probability density of the initial magnetization $m_0(t)$, conditioned on relaxing to vanishing magnetization $m=0$ at time $t$. The initial magnetization $m_0(t)$ as a function of time serves as an order parameter for the finite-time dynamical phase transition, analogous to the magnetization $m$ as a function of $\mathscr{J}$ at equilibrium. 

Figure~\ref{fig:rates_mf}(c) shows the probability density of the absolute value $|m_0|$ of the initial magnetizations of relaxation trajectories that reach $m=0$ at different times $t$ after the quench. The heat map shows the normalized density, with yellow again corresponding to the highest density. The solid line is obtained from the mean-field calculation~\cite{jan_massi_prl}, valid as $L\to\infty$. The dashed line marks the critical time, Eq.~\eqref{eq:tc_mf}. 

We note the striking similarity between the behaviors of $|m_0|$ as a function of $t$ in Fig.~\ref{fig:rates_mf}(c) and of the magnetization $|m|$ as a function of $\mathscr{J}$ at equilibrium, shown in Fig.~\ref{fig:equilibrium}(a). Importantly, for the finite-time dynamical phase transition time $t$ takes the role of the control parameter $\mathscr{J}$ at equilibrium~\cite{jan_massi_prl}.
\section{Results}\label{sec:results}
The existence of finite-time dynamical phase transitions has been established in both the mean-field model~\cite{jan_massi_prl,jan_massi_njp} and the NNIM~\cite{BlomGodec2023}. We now analyze the statistics of the initial magnetization $m_0$ (the order parameter) in the vicinity of $t_c$, i.e., close to the critical point, to gain insight into the critical phenomena associated with the finite-time dynamical phase transition in the NNIM.

To this end, we quench a large but finite system from the ordered (ferromagnetic) phase with $\mathscr{J}>\mathscr{J}_c$ into the disordered (paramagnetic) phase where $\mathscr{J}_q<\mathscr{J}_c$, and evolve it numerically with the spin-flip dynamics discussed in Sec.~\ref{sec:dynamics}. To obtain the time-dependent rate function $V(m,t)$ from Eq.~\eqref{eq:rfneq}, we record the evolution of the system and calculate $P(m,t)$ from a large number (a minimum of $10^7$) of realizations of the relaxation process.
\subsection{Finite-time dynamical phase transition}
\begin{figure*}[htp]
\centering
  \includegraphics*[width=\linewidth]{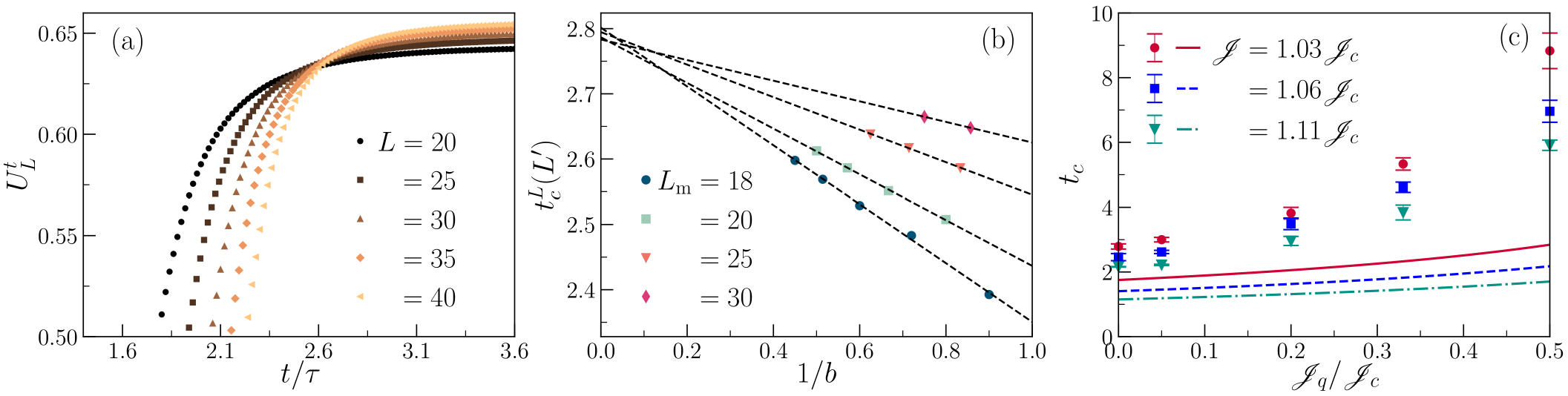}
  \caption{Determination of the critical time from a series of finite-size measurements. (a) Dynamical Binder cumulant, $U_L^t$~\eqref{eq:variance}, as function of $t/\tau$ for different system sizes for a quench from $\mathscr{J} = 1.03 \mathscr{J}_c$ to $\mathscr{J}_q = 0$. (b) System-size dependent critical time $t_c^L(L')$ for $\mathscr{J} = 1.03 \mathscr{J}_c$ and $\mathscr{J}_q = 0$, obtained from intersections of Binder cumulants for various series of system sizes -- see text. The dashed lines show linear fits used to calculate $t_c$. (c) Critical times for quenches from three different initial $\mathscr{J}$ values to different final points $\mathscr{J}_q$. The lines are from mean-field calculations~\cite{jan_massi_prl}.}
  \label{fig:ul}
\end{figure*}
Figure~\ref{fig:rates_nn}(a) shows the time-dependent rate function $V(m,t)$ of the NNIM, as a function of $m$ at different times after the quench as the colored symbols. The vertical dashed lines mark the positions of the minima at $\pm m_\text{eq}$ in the infinite system size limit, given in Eq.~\eqref{eq:meqons}. Similar to the rate function of the mean-field model, shown in Fig.~\ref{fig:rates_mf}(a), the rate function of the NNIM forms a kink at $m=0$ during the relaxation. However, due to the finite-size nature of the simulations, the kink is not quite sharp.

More conclusive evidence for the finite-time dynamical phase transitions is obtained by considering the optimal fluctuations that achieve vanishing magnetization, $m=0$, at different times. Figure~\ref{fig:rates_nn}(b) shows the optimal fluctuations that realize $m=0$ at short and longer times in terms of their trajectory densities at finite system size. The densities of conditioned relaxation trajectories are shown in terms of a heat map with yellow corresponding to regions of maximum density. The dashed line marks the critical time $t^L_c\approx 2.4\tau$ obtained numerically using a method explained in Sec.~\ref{sec:tc}. We observe a characteristic difference between the optimal fluctuations for times shorter and longer than the critical time, very similar to the mean-field case [see Fig.~\ref{fig:rates_mf}(b)]: For short times, the optimal fluctuation initiates at an unlikely initial magnetization close to $m=0$ and stays there, while for longer times it initiates close to the (most likely) spontaneous magnetization $\pm m_\text{eq}$ shown as the horizontal dash-dotted lines in Fig.~\ref{fig:rates_nn}(b).

As in the mean-field version of the model, the occurrence of competing relaxation behaviors at short and longer times suggests that at the finite critical time a transition between the two takes place, characterized by the distribution of their initial magnetizations $m_0(t)$. Figure~\ref{fig:rates_nn}(c) shows the probability density of $|m_0|$ of relaxation trajectories that reach $m=0$ at time $t$. 
Indeed, we observe that $m_0$ is essentially zero at short times, but assumes finite values at times longer than $t_c^L$. The behavior is analogous to that of $m_0(t)$ in the mean-field model shown in Fig.~\ref{fig:rates_mf}(c) and to the statistics of $|m|$ at equilibrium, see Fig.~\ref{fig:equilibrium}(a).
\subsection{Critical time}\label{sec:tc}
Motivated by the striking similarities between the finite-time dynamical phase transition and the equilibrium phase transition [see Figs.~\ref{fig:equilibrium}(a), \ref{fig:rates_mf}(c), and \ref{fig:rates_nn}(c)], we apply equilibrium methods to analyze the finite-time dynamical phase transition in the vicinity of the critical point $t=t_c$.

For equilibrium phase transitions, a common way to obtain critical points is by evaluating the so-called Binder cumulant of the order-parameter distribution~\cite{binder1981finite,binder1981critical}, defined as $U_L^{\rm{eq}} = 1 -\langle m_\text{eq}^4\rangle_L/(3\langle m_\text{eq}^2 \rangle_L ^2)$. In the limit $L \rightarrow \infty$, the Binder cumulant takes, respectively, the values $0$ and $2/3$ in the paramagnetic phase and the ferromagnetic phase. At the critical point, $U_L^{\rm{eq}}$ is approximately independent of the system size. Therefore, the intersection of different Binder cumulants calculated at different system sizes provides a measure for the location of the critical point from a series of finite-size measurements.

We adopt this method to calculate the critical time $t_c$ of the finite-time dynamical phase transition in the NNIM. To this end, we define a dynamical Binder cumulant as
\eqn{\label{eq:variance}
       U_L^t = 1 - \frac{\langle m_0^4\rangle_L}{3\langle m_0^2 \rangle_L ^2}.
}
Figure~\ref{fig:ul}(a) shows $U_L^t$ as a function of $t/\tau$. We observe that the Binder cumulants for different system sizes intersect not quite at the same point, but in a small range of $t/\tau$ values. To infer the intersection point of $U_L^t$ in the thermodynamic limit $L\to\infty$, we compute the intersection of $U_L^t$ for increasing series of system sizes, thus providing estimates of the system-size dependent critical time, $t_c^L$.

More precisely, for each system size $L_m$, we take a series of larger systems of sizes $L'$, differing by a scaling factor $b>1$, i.e., $L' = b L_m$, so that $L_m$ corresponds to the smallest system in each series. For each of these series, we compute the intersections of the Binder cumulants as a function of $b$. Figure~\ref{fig:ul}(b) shows these intersections for different $L_m$ with all larger $L'>L_m$, as a function of the inverse of the scale factor, $1/b$.

Extrapolation of the data to $1/b \rightarrow 0$ by weighted linear fits (dashed lines) then provides an estimate of the critical time in the thermodynamic limit $L\to\infty$. For $\mathscr{J} = 1.03\mathscr{J}_c$ and $\mathscr{J}_q=0$, the average over all such extrapolations gives $t_c=2.79 \pm 0.08$. The values for $t_c$ obtained in this way are consistent with those from other scaling methods, based on, e.g., different extrapolation schemes~\cite{henkel1988finite,west2015efficient}.

Applying the method illustrated in Fig.~\ref{fig:ul}(b) at different inverse coupling temperatures $\mathscr{J}_q$, we obtain $t_c$ as a function of $\mathscr{J}_q$ for three different $\mathscr{J}$ values. Figure~\ref{fig:ul}(c) shows the result of this analysis, including error bars representing 95\,\% confidence intervals, estimated using a jackknife resampling method~\cite{bootstrap}. The numerical values of $t_c$ are tabulated in Tab.~\ref{tab:tc} in Appendix~\ref{sec:App_tc}.

The lines in Fig.~\ref{fig:ul}(c) show the exact results for the mean-field model given in Eq.~\eqref{eq:tc_mf}~\cite{ermolaev2010low,jan_massi_prl}. We observe that while the critical time of the mean-field model is consistently smaller than that of the NNIM, $t_c$ is an increasing function of $\mathscr{J}_q$ and a decreasing function of $\mathscr{J}$ in all cases.
\subsection{Generalized susceptibility}\label{sec:chi0}
After having obtained an estimate of the critical time, we now explore the critical properties of the transition, adopting the finite-size scaling analysis reviewed in Sec.~\ref{sec:fss}.

\begin{figure}[htp]
\centering
  \includegraphics*[width=0.4\textwidth]{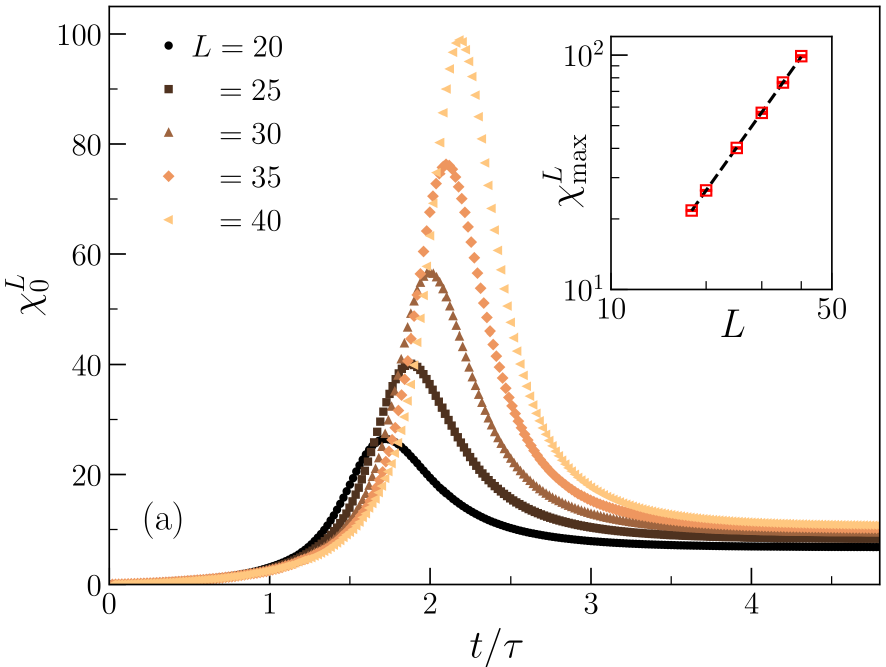}
   \vskip 0.3cm
  \includegraphics*[width=0.4\textwidth]{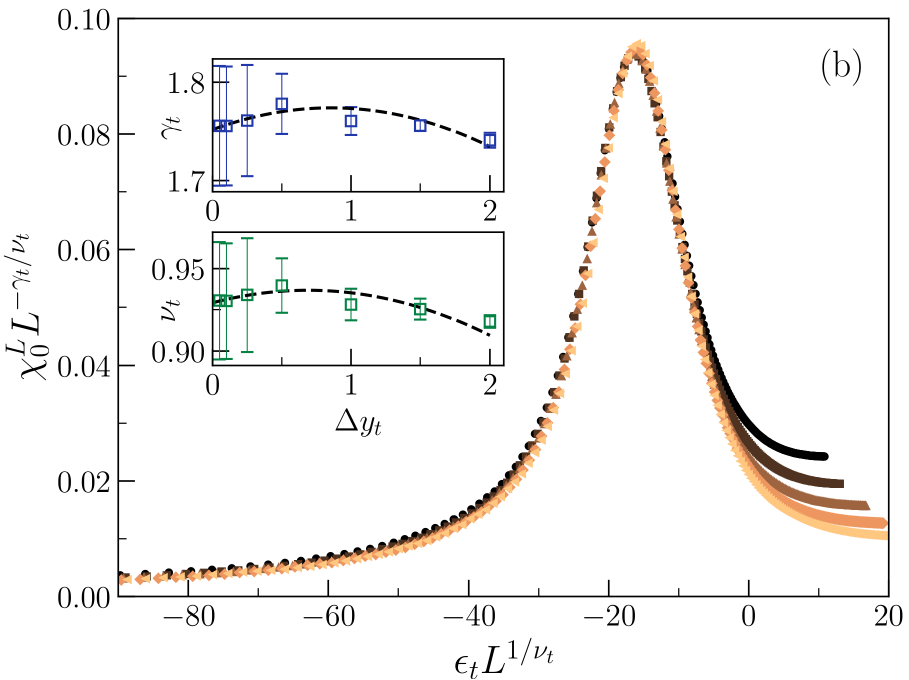} 
  \caption{Finite-size scaling of the generalized susceptibility $\chi_0^L$ defined in Eq.~\eqref{eq:chi0} for a quench from $\mathscr{J} = 1.03 \mathscr{J}_c$ to $\mathscr{J}_q = 0$. (a) $\chi_0^L$ as a function of $t/\tau$ for different system sizes. Inset: $\chi_{\rm{max}}$ as a function of $L$ on a log-log scale. The dashed line is a power-law fit with exponent $ 1.89 \pm 0.01$ ($ = \gamma_t/\nu_t$). The errors are within the symbol size. 
  (b) Collapse of $\chi_0^L$ from different system sizes onto the scaling function $Y_{\chi_0^L}$ using the parameters $t_c = 2.84$, $\gamma_t = 1.75$, and $\nu_t = 0.93$. Insets: Estimates of $\gamma_t$ (top) and $\nu_t$ (bottom) from minimizing $\sigma_{Y_{\chi_0}}$ in Eq.~\eqref{eq:sigx} for different intervals $\Delta y_t$. The dashed lines are quadratic fits to the data.}
  \label{fig:chi}
\end{figure}

We define the time-dependent, generalized susceptibility $\chi_0^L$ of the initial magnetization $m_0$ at finite system size, the non-equilibrium version of Eq.~\eqref{eq:susc}, as
\eqn{\label{eq:chi0}
        \chi_0^L := L^{d} (\langle m_0^2 \rangle - \langle |m_0| \rangle^2)\,.
}
In the non-equilibrium case, time plays the role of the inverse coupling temperature at equilibrium. We therefore define
\eqn{
    \epsilon_t := \frac{t-t_c}t\,,
    }
analogously to $\epsilon$ in Eq.~\eqref{eq:beta_eq}. As in the case of the equilibrium transition, in the thermodynamic limit, the susceptibility $\chi_0$ of the infinite system diverges at the critical point. Analogous to the equilibrium case, we expect the scaling of $\chi_0$ as a function of $\epsilon_t$ to be
\eqn{\label{eq:chi_neq}
       \chi_0 \sim |\epsilon_t|^{-\gamma_t}\,,
}
where $\gamma_t$ denotes the critical exponent associated with the generalized susceptibility. For finite systems, Eq.~\eqref{eq:chi_neq} needs to be modified as $\chi_0^L(t_c) \sim L^{\gamma_t/\nu_t}$, cf. Eq.~\eqref{eq:mag_eq_scaling}.

The exponent $\nu_t$ is the critical exponent associated with the time-dependent correlation length, $\xi_t$. Analogous to Eq.~\eqref{eq:chi_eq_scaling}, we define a dynamical scaling function as,
\eqn{\label{eq:chi_scaling}
Y_{\chi_0}(y_t) := \chi_0^L L^{-\gamma_t/\nu_t},
}
where 
\eqn{\label{eq:scalvar_neq}
	y_t  :=  \epsilon_t L^{1/\nu_t} \sim \text{sign}(\epsilon_t)\left(\frac{L}{\xi_t}\right)^{1/\nu_t}\,,
}
is the time-dependent scaling variable.

Figure~\ref{fig:chi}(a) shows $\chi_0^L$ for different system sizes. Analogous to $\chi_\text{eq}$ in Fig.~\ref{fig:equilibrium}(b), $\chi_0^L$ shows a pronounced peak whose magnitude becomes larger as the system size increases.

In perhaps the most direct way, $\gamma_t/\nu_t$ can be estimated from the maximum of the susceptibility, $\chi_{\rm{max}}^L$ for different system sizes. With the system size, $\chi_{\rm{max}}^L$ scales as~\cite{privman1990}
\eqn{\label{eq:chimax_scaling}
    \chi_{\rm{max}}^L \sim L^{\gamma_t/\nu_t}.
}
The inset of Fig.~\ref{fig:chi}(a) confirms the power-law scaling of $\chi_{\rm{max}}^L$ as a function of $L$. Together with Eq.~\eqref{eq:chimax_scaling}, a power-law fit in principle provides an estimate for $\gamma_t/\nu_t$. In practice, however, this method is error prone, since it depends on $\chi_0^L$ evaluated at a single point (the maximum $\chi_{\rm{max}}^L$). A more accurate method relies on determining the exponents from finite intervals $\Delta y_t$ of the dynamical scaling function $Y_{\chi_0}(y_t)$~\cite{newman1996monte} in Eq.~\eqref{eq:chi_scaling}, and extrapolating to the limit $\Delta y_t\to 0$.

An accurate treatment of finite-size effects in the non-equilibrium data turns out to be crucial, because these effects arise from multiple sources: the initial configuration, the finite-time conditioning, and the relaxation dynamics.  In particular, the initial inverse coupling temperature, $\mathscr{J}$, as well as the initial equilibrium rate function $V_\text{eq}$ are affected by finite system size because $\mathscr{J}$ is close to the critical point $\mathscr{J}_c$ at equilibrium. However, initiating the system close to $\mathscr{J}_c$ is required to obtain a sufficiently large ensemble of conditioned relaxation trajectories. In addition, the dynamics is also affected by finite $L$, because first, $L$ needs to be small enough to ensure a sufficient sample size and second, one needs to choose a small (but large enough) window around $m=0$ (of order $\sim N^{-\frac12}$ here) for the conditioning on $m=0$ at different $t$. Finite-size effects cumulating from these sources explain why a careful finite-size scaling analysis is required to determine the critical exponents.

We determine the parameters $\gamma_t$, $\nu_t$, and $t_c$ by optimizing the data in Fig.~\ref{fig:chi}(a) to achieve the best possible collapse onto the scaling function $Y_{\chi_0}$. To this end, we minimize the variance $\sigma_{Y_{\chi_0}}$ of the data sets from different system sizes obtained by integrating over a range of scaling variables $y_t\in[y^{\rm{min}}_t,y^{\rm{max}}_t]$ near the critical point~\cite{newman,newman1996monte}:
\eqn{\label{eq:sigx}
    \sigma_{Y_x}^2 = \frac{1}{\Delta y_t}\int_{y^{\rm{min}}_t}^{y^{\rm{max}}_t}\!\!dy_t\, \bigg\{N_L \sum_L [Y_x (y_t)]^2 - \Big[\sum_L Y_x(y_t)\Big]^2 \bigg\}\,,
}
where $\Delta y_t := y^{\rm{max}}_t-y^{\rm{min}}_t$ is a variable $y_t$ interval, $N_L$ is the number of system sizes considered, and $Y_x(y_t)$ is the scaling function of either $x = \chi_0$ or $x = m_0$ (defined in Sec.~\ref{sec:m0}).

By minimizing $\sigma_{Y_{\chi_0}}$ in this way, we obtain estimates for $t_c$ and the critical exponents $\gamma_t$ and $\nu_t$ for $\mathscr{J} = 1.03 \mathscr{J}_c$ and $\mathscr{J}_q = 0$. Figure~\ref{fig:chi}(b) shows the collapse for the optimal values, $t_c =2.84\pm 0.03$, $\gamma_t= 1.75\pm0.02$, and $\nu_t= 0.93\pm0.01$. The insets show how $\gamma_t$ and $\nu_t$ are estimated from Eq.~\eqref{eq:sigx} with varying intervals $\Delta y_t$~\cite{newman1996monte}. A quadratic fit to the data gives our best estimates for $\gamma_t$ and $\nu_t$ in the limit $\Delta y_t \rightarrow 0$, reported in Sec.~\ref{sec:critexp}.
\begin{figure}[htp]
\centering
  \includegraphics*[width=0.4\textwidth]{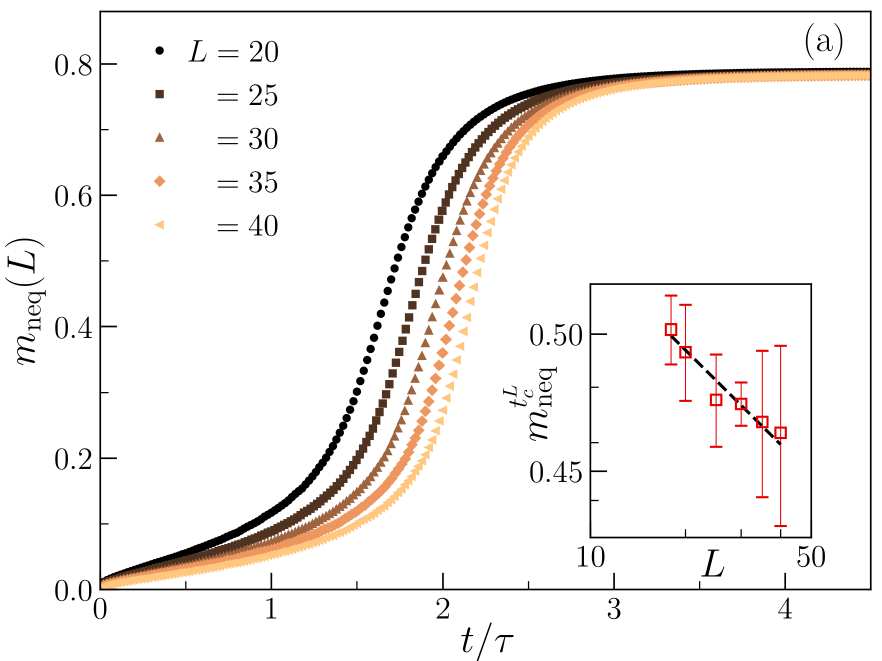}
  \vskip 0.3cm
  \includegraphics*[width=0.4\textwidth]{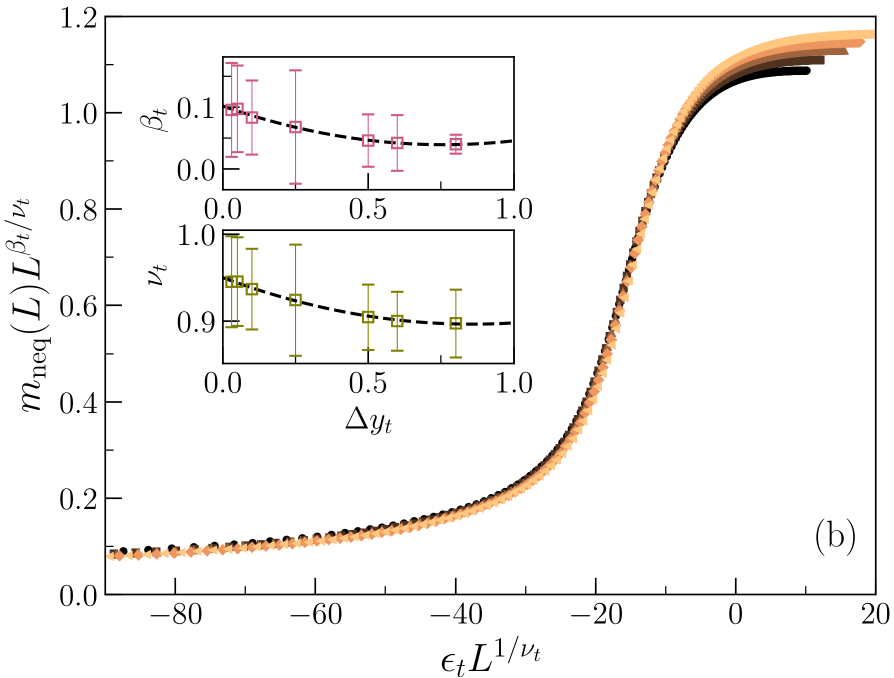} 
  \caption{Finite-size scaling of the spontaneous initial magnetization $m_\text{neq}(L)$ in Eq.~\eqref{eq:mneq} for a quench from $\mathscr{J} = 1.03 \mathscr{J}_c$ to $\mathscr{J}_q = 0$. (a) $m_\text{neq}(L)$ as a function of $t/\tau$ for different system sizes. Inset: $m_{\text{neq}}^{t_c^L}$, as a function of $L$ on log-log scale (see text). The dashed line is a power-law fit with exponent $0.105$ ($=\beta_t/\nu_t$). (b) Collapse of $m_\text{neq}(L)$ onto the scaling function $Y_{m_0}$~\eqref{eq:mag_t_scaling} using the parameters $t_c = 2.84$, $\beta_t = 0.102$, and $\nu_t = 0.95$. Insets: Estimates of $\beta_t$ (top) and $\nu_t$ (bottom) from minimizing $\sigma_{Y_{m_0}}$ in Eq.~\eqref{eq:sigx} for varying intervals $\Delta y_t$. The dashed lines are quadratic fits to the data.}\label{fig:betat}
\end{figure}

\subsection{Initial magnetization}\label{sec:m0}
In a similar way, we extract the critical exponents associated with the initial magnetization, $m_0$. In the limit $L\to\infty$, the value of the spontaneous initial magnetization
\eqn{\label{eq:mneq}
    m_\text{neq} := \langle |m_0|\rangle\,,   
}
vanishes at $t<t_c$ but becomes finite for $t>t_c$, analogous to $m_\text{eq}$ as a function of $\mathscr{J}$ in the equilibrium case. At finite $L$, $m_\text{neq}(L)$ is expected to show a similar behavior, although not as sharp. Figure~\ref{fig:betat}(a) shows $m_\text{neq}(L)$ for a quench from $\mathscr{J} = 1.03 \mathscr{J}_c$ to $\mathscr{J}_q = 0$ and for different system sizes. We observe that as $L$ increases, $m_\text{neq}(L)$ exhibits an increasingly sharp transition to finite values around the critical time $t_c\approx 2.8\tau$. In the thermodynamic limit we expect the scaling
\eqn{
    m_\text{neq} \sim \epsilon_t^{\beta_t}\,,
}
with critical exponent $\beta_t$ for the spontaneous initial magnetization, while approaching from above $t_c$, analogous to the equilibrium case, cf. Eqs.~\eqref{eq:mag_eq_scaling}. 

For finite $L$, we should instead have
\eqn{\label{eq:mneqL}
	m_\text{neq}(L) \sim L^{\beta_t/\nu_t}\,,
}
at the size-dependent critical time $t = t_c^L$, obtained with the method in Sec~\ref{sec:tc}. The inset in Fig.~\ref{fig:betat}(a) shows that the data for $m_\text{neq}(L)$ at $t_c^L$, denoted as $m_\text{neq}^{t_c^L}$ are consistent with the power-law scaling~\eqref{eq:mneqL} for the ratio $\beta_t/\nu_t =  0.105 \pm 0. 025$.

A scaling-function analysis similar to that in Sec.~\ref{sec:chi0} provides a more accurate estimate of $\beta_t$. We define the analog of Eq.~\eqref{eq:beta_eq_scaling}, the scaling function $Y_{m_0}$, as
\eqn{\label{eq:mag_t_scaling}
Y_{m_0}(y_t) := m_\text{neq}(L) L^{\beta_t/\nu_t}\,,
}
with $y_t$ defined in Eq.~\eqref{eq:scalvar_neq}. 

To obtain estimates for $\beta_t$ and $\nu_t$ based on Eq.~\eqref{eq:mag_t_scaling}, we minimize the variance $\sigma_{Y_{m_0}}$ in Eq.~\eqref{eq:sigx} to obtain the best possible collapse. To make the minimization algorithm converge, we fix $t_c$ to the values determined by the scaling in Sec.~\ref{sec:chi0}. The insets in Fig.~\ref{fig:betat}(b) show the result of this minimization, yielding $\beta_t = 0.102 \pm 0.056$ (top) and $\nu_t = 0.95 \pm 0.03$ (bottom). The main panel of Fig.~\ref{fig:betat}(b) shows how $m_\text{neq}(L)$ collapses onto $Y_{m_0}$ for these values of $\beta_t$ and $\nu_t$.
\subsection{Critical exponents}\label{sec:critexp}
\begin{figure}[htp]
\centering
  \includegraphics*[width=\linewidth]{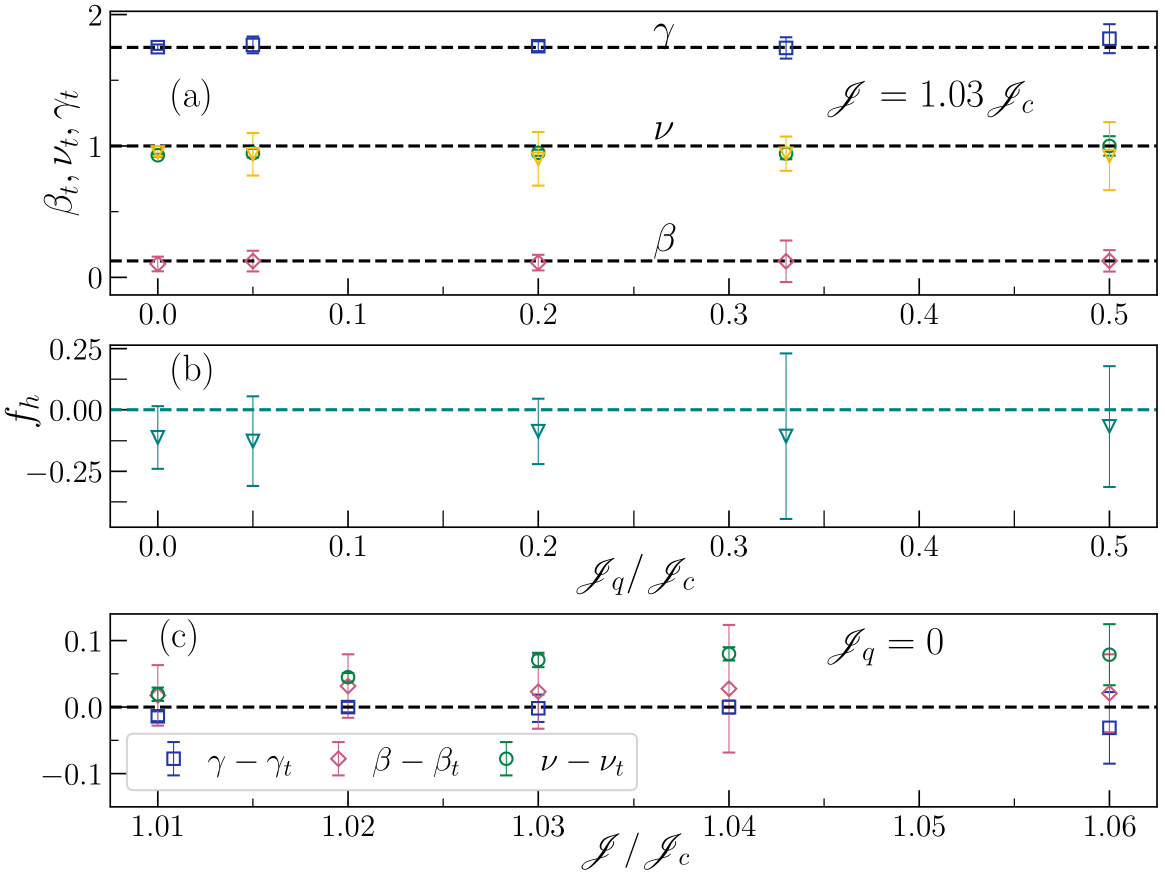}
  \caption{(a) Critical exponents, $\beta_t$, $\gamma_t$, and $\nu_t$ for different $\mathscr{J}_q$. Dashed horizontal lines correspond to the exponents in the two dimensional Ising universality class. (b) $f_h$ for different $\mathscr{J}_q$, showcasing consistency of the exponents in (a) with the hyperscaling relation~\eqref{eq:hyperscaling} (see text). (c) Differences between the dynamical and equilibrium critical exponents as functions of $\mathscr{J}/\mathscr{J}_c$ for fixed $\mathscr{J}_q (= 0)$.}
  \label{fig:exponents}
\end{figure}
Repeating the analysis summarized in Secs.~\ref{sec:chi0} and \ref{sec:m0} for different quenched inverse coupling temperatures, $\mathscr{J}_q$, we obtain the critical exponents $\beta_t$, $\gamma_t$, and $\nu_t$ as functions of $\mathscr{J}_q$. Figure~\ref{fig:exponents}(a) shows the result of this analysis for $\mathscr{J} = 1.03 \mathscr{J}_c$. The dashed lines show the corresponding equilibrium exponents of the two dimensional Ising universality class. For the critical exponent $\nu_t$, we obtain two estimates, shown as green circles and yellow triangles, that are obtained from the analyses in Secs.~\ref{sec:chi0} and \ref{sec:m0}, respectively. The values for $\gamma_t$ and $\beta_t$ are obtained from minimizing $\sigma_{Y_{\chi_0}}$ (Sec.~\ref{sec:chi0}, blue squares) and $\sigma_{Y_{m_0}}$ (Sec.~\ref{sec:m0}, red diamonds), respectively. We observe that although the critical time $t_c$ depends on both the initial and final point of the quenching process, $\mathscr{J}$ and $\mathscr{J}_q$, respectively [see Fig.~\ref{fig:ul}(c)], the critical exponents remain constant for varying $\mathscr{J}_q$. Furthermore, the value of $\gamma_t$ is consistent with the value $\gamma=\frac{7}{4}$ of the two-dimensional Ising universality class at equilibrium. However, the value of $\nu_t$ is consistently \textit{smaller} than the equilibrium value $\nu=1$. Finally, $\beta_t$, although consistent with the equilibrium value $\beta=\frac{1}{8}$, carries a comparably large error. The numerical values of the calculated exponents are listed in Tab.~\ref{tab:exp_J1.03Jc} in Appendix~\ref{sec:App_tc}.

The apparent consistency of $\beta_t$ and $\gamma_t$ with the equilibrium values $\beta$ and $\gamma$, and the fact that $\nu_t$ is consistently smaller than $\nu$, pose the question of whether the so-called hyperscaling relation~\cite{fisher_scaling}
\eqn{\label{eq:hyperscaling}
	\nu_t d = 2\beta_t + \gamma_t\,,
}
holds, where $d=2$ denotes the spatial dimension. Hyperscaling connects all critical exponents and it arises whenever the correlation length $\xi_t$ is the only relevant length scale in the infinite system at the critical point~\cite{privman1990}. Since the latter property forms the basis of our scaling analysis, and due to the analogies between the finite-time dynamical phase transition and the equilibrium transition, we expect~\eqref{eq:hyperscaling} to hold.

This is confirmed in Fig.~\ref{fig:exponents}(b), which shows $f_h = \nu_t d - 2\beta_t - \gamma_t$ as function of $\mathscr{J}_q/\mathscr{J}_c$. In all considered cases, the measured values of $f_h$ are consistent with hyperscaling~\eqref{eq:hyperscaling}, i.e., $f_h=0$, within the error bars. However, because of the uncertainties in the values for $\beta_t$, the error for $f_h$ is quite large.

Assuming that hyperscaling~\eqref{eq:hyperscaling} holds, a logical consequence of the fact that $\nu_t$ departs from the equilibrium value $\nu$ is that either one or both of the remaining exponents $\beta_t$ and $\gamma_t$ must deviate from their equilibrium counter parts. However, the sizes of the errors of $\gamma_t$ and in particular of $\beta_t$ currently do not permit us to resolve these deviations.

Finally, we analyze the dependence of the critical exponents on the initial inverse coupling temperature $\mathscr{J}$. To this end, we repeat the scaling analysis summarised in Secs.~\ref{sec:chi0} and \ref{sec:m0} for different $\mathscr{J}$ but at fixed quenched inverse coupling temperature $\mathscr{J}_q = 0$.

Figure~\ref{fig:exponents}(c) shows the differences between the values of the critical exponents and the corresponding values from the two-dimensional Ising universality class as functions of $\mathscr{J}/\mathscr{J}_c$. We observe that $\gamma_t$ remains at the equilibrium value $\gamma=\frac{7}{4}$ for all measured $\mathscr{J}$. A similar conclusion holds for $\beta_t$, but in this case the errors are simply too large to identify small deviations from $\beta=\frac{1}{8}$. The exponent $\nu_t$, by contrast, is consistently smaller than $\nu=1$ for all measured $\mathscr{J}$, but approaches $\nu$ as the initial system approaches the equilibrium critical point, i.e., $\mathscr{J}\to\mathscr{J}_c$. The numerical values for the exponents are listed in Tab.~\ref{tab:exp_Jq0} in Appendix~\ref{sec:App_tc}.

In other words, as the initial inverse coupling temperature $\mathscr{J}$ approaches the equilibrium critical point $\mathscr{J}_c$, the values of the dynamical critical exponents $\beta_t$, $\gamma_t$, and $\nu_t$ approach the exponents $\beta$, $\gamma$ and $\nu$ of the corresponding equilibrium phase transition, across which the system is initially quenched, just as for the mean-field model \cite{jan_massi_prl}. Away from the critical point, the values of $\beta_t$ and $\gamma_t$ are both consistent with $\beta$ and $\gamma$, but $\nu_t$ is consistently smaller than $\nu$. Meanwhile, the hyperscaling relation~\eqref{eq:hyperscaling} remains fulfilled within the error bars. Hyperscaling and the deviations of $\nu_t$ from $\nu$ imply that there must be additional deviances of either $\beta_t$ and/or $\gamma_t$ from their equilibrium counter parts, which, however, we cannot resolve with the present method.
\section{Conclusions}\label{sec:conclusion}
We have studied critical phenomena associated with the finite-time dynamical phase transition of the nearest-neighbor Ising model on a two-dimensional square lattice by means of Monte Carlo simulations. The transition manifests itself as a finite-time switch in the most likely relaxation dynamics, the optimal fluctuation, after a disordering quench from the ferromagnetic phase into the paramagnetic phase~\cite{jan_massi_prl,jan_massi_njp,BlomGodec2023}. The initial magnetization $m_0(t)$ associated with the optimal fluctuation of the relaxation dynamics that achieves $m=0$ at given time $t$ serves as an order parameter for the transition, similar to the magnetization at equilibrium.

Using a finite-size scaling analysis, we extracted both the critical time $t_c$ at which the transition occurs, as well as the associated critical exponents from the statistics of $m_0$. Near the critical point, the fluctuations of $m_0$, measured by the generalized susceptibility $\chi_0$, diverge and exhibit characteristics of a continuous phase transition. We explored the critical fluctuations of $m_0$ using the scaling functions $Y_{m_0}$ and $Y_{\chi_0}$ associated with $m_0$ and $\chi_0$, respectively. From this analysis, we estimated the dynamical critical exponents $\beta_t$, $\gamma_t$, and $\nu_t$ corresponding to the initial magnetization $m_0$, the generalized susceptibility $\chi_0$, and the correlation length $\xi_t$ of $m_0$, respectively.

When the system is initially in the vicinity of the equilibrium critical point, the obtained values of these exponents were found to be consistent with the values in the equilibrium counter part of the model. For initial inverse coupling temperatures $\mathscr{J}$ away from the critical point, however, the exponent $\nu_t$ is smaller than that of the equilibrium value $\nu$. All exponents are independent of the quenched inverse coupling temperature $\mathscr{J}_q$, and consistent with the hyperscaling relation~\eqref{eq:hyperscaling}. Hyperscaling further implies that one or two of the remaining exponents $\beta_t$ and $\gamma_t$ must be different from their equilibrium counter parts. Unfortunately, we were unable to resolve these additional deviations with the current method.

We conclude that when the system is initially in the vicinity of the equilibrium critical point, the finite-time dynamical phase transition in the nearest-neighbor Ising model belongs to the same universality class as the corresponding equilibrium model, consistent with the results from the mean-field model~\cite{jan_massi_prl}, where this conclusion holds for all $\mathscr{J}$. For $\mathscr{J}$ values away from the critical point, the observed deviations of the exponent $\nu_t$ from the equilibrium value indicate that finite-time dynamical phase transitions are genuine non-equilibrium phenomena whose critical properties are generally distinct from their equilibrium counter parts.

The observed dependencies of the exponent $\nu_t$ and the critical time $t_c$ on the initial coupling temperature $\mathscr{J}$ of the quench suggest that correlations between spins in the initial states of the transition play an important role. In the future, we intend to analyze the spatial structure of the Ising model during the relaxation process achieving $m=0$ at a given time $t$ to obtain a clearer understanding of, e.g., the monotonous dependence $t_c$ on the external parameters.

It would furthermore be interesting to investigate the finite-time dynamical phase transition in higher dimensions $d>2$. This would clarify whether or not the dynamical critical exponents, quenched from the critical point, agree with the equilibrium one also in this case. Crucially, at equilibrium the nearest-neighbor Ising model becomes mean-field like for $d\geq4$, and it is a fascinating open question if this behavior is paralleled in the non-equilibrium transition. The use of phenomenological theories such as Ginzburg-Landau theory~\cite{bray,hohenberg1977} could perhaps be useful to elucidate these questions.

\begin{acknowledgments}
This research was funded by the Project INTER/FNRS/20/15074473 ``TheCirco'' on Thermodynamics of Circuits for Computation, funded by the F.R.S.-FNRS (Belgium) and FNR (Luxembourg). The simulations were carried out using the HPC facilities of the University of Luxembourg \cite{VCPKVO_HPCCT22} {\small -- see \url{https://hpc.uni.lu}}. JM's stay at King's College London was funded by a Feodor-Lynen fellowship of the Alexander von Humboldt-Foundation.
\end{acknowledgments}
\begin{widetext}
\appendix
 \section{Critical time and critical exponents} \label{sec:App_tc}
In this Appendix, we summarize the numerical values for the critical time and for the critical exponents, obtained with the methods described in the main text. The associated errors represent 95\,\% confidence intervals, estimated using a jackknife resampling method~\cite{bootstrap}.

In Tab.~\ref{tab:tc} we list the values of the critical time calculated with the method described in Sec.~\ref{sec:tc}.

\begin{table}[htp]
\caption{Critical time, $t_c$, calculated for different values of $\mathscr{J}$ and $\mathscr{J}_q$.}\label{tab:tc}
\begin{tabular}{c|c|c|c|c|c}
\hline
\multirow{2}{*}{$\frac{\mathscr{J}}{\mathscr{J}_c}$} & \multicolumn{5}{|c}{$\mathscr{J}_q/\mathscr{J}_c$}\\
\cline{2-6}
& {$0$} &{$0.05$} & {$0.2$} & {$0.33$} & {$0.5$}\\
\hline
{$1.03$} & {$2.79\pm 0.08$} & {$3.00\pm 0.07$}& {$3.82\pm 0.18$} & {$5.33\pm 0.19$} & {$8.83\pm 0.55$} \\
\hline
{$1.06$} & {$2.46\pm 0.11$} & {$2.62\pm 0.05$} & {$3.49\pm 0.18$} & {$4.61\pm 0.16$} & {$6.96\pm 0.34$}\\
\hline
{$1.11$} & {$2.16\pm 0.01$} & {$2.22\pm 0.02$} & {$2.96\pm 0.14$} & {$3.84\pm 0.23$} & {$5.91 \pm 0.16$} \\
\hline
\end{tabular}
\end{table}
Table~\ref{tab:exp_J1.03Jc} gives the values of the critical exponents $x_t=\{\gamma_t,\beta_t,\nu_t^{(1)},\nu_t^{(2)}\}$ calculated from the finite-size scaling analysis of $\chi_0$ and $m_{\rm{neq}}$ for quenches from fixed $\mathscr{J} = 1.03\mathscr{J}_c$ to different $\mathscr{J}_q$. The values for $\nu_t^{(1)}$ and $\nu_t^{(2)}$ are calculated from the scaling of $\chi_0$ and $m_{\rm{neq}}$, respectively.

In Tab.~\ref{tab:exp_Jq0} we list the same exponents $x_t$ for quenches from different $\mathscr{J}$ to a fixed $\mathscr{J}_q = 0$.
\begin{table*}[htp]
\caption{Critical exponents, $x_t=\{\gamma_t,\beta_t,\nu_t^{(1)},\nu_t^{(2)}\}$, calculated for quenches from a fixed value of $\mathscr{J} (= 1.03\mathscr{J}_c)$ to different $\mathscr{J}_q$ values.}\label{tab:exp_J1.03Jc}
\begin{tabular}{c|c|c|c|c|c}
\hline
\multirow{2}{*}{$x_t$} & \multicolumn{5}{|c}{$\mathscr{J}_q/\mathscr{J}_c$}\\
\cline{2-6}
& {$0$} &{$0.05$} & {$0.2$} & {$0.33$} & {$0.5$}\\
\hline
{$\gamma_t$} & {$1.75\pm 0.02$} & {$1.77\pm 0.06$}& {$1.76\pm 0.04$} & {$1.75 \pm 0.08$} & {$1.82\pm 0.11$} \\
\hline
{$\beta_t$} & {$0.102\pm 0.056$} & {$0.124\pm 0.079$} & {$0.112\pm 0.059$} & {$0.122\pm 0.158$} & {$0.126\pm 0.081$}\\
\hline
{$\nu_t^{(1)}$} & {$0.93\pm 0.01$} & {$0.94 \pm 0.03$} & {$0.95\pm 0.02$} & {$0.94\pm 0.04$} & {$1.0 \pm 0.07$} \\
\hline
{$\nu_t^{(2)}$} & {$0.95\pm 0.03$} & {$0.94 \pm 0.16$} & {$0.9\pm 0.2$} & {$0.94\pm 0.13$} & {$0.94 \pm 0.26$} \\
\hline
\end{tabular}
\end{table*}
\begin{table*}[htp]
\caption{Critical exponents, $x_t=\{\gamma_t,\beta_t,\nu_t^{(1)},\nu_t^{(2)}\}$, calculated for quenches from different initial values of $\mathscr{J}$ to a fixed $\mathscr{J}_q (=0)$.}\label{tab:exp_Jq0}
\begin{tabular}{c|c|c|c|c|c}
\hline
\multirow{2}{*}{$x_t$} & \multicolumn{5}{|c}{$\mathscr{J}/\mathscr{J}_c$}\\
\cline{2-6}
& {$1.01$} &{$1.02$} & {$1.03$} & {$1.04$} & {$1.06$}\\
\hline
{$\gamma_t$} & {$1.76\pm 0.01$} & {$1.75\pm 0.01$}& {$1.75\pm 0.02$} & {$1.75 \pm 0.01$} & {$1.78\pm 0.05$} \\
\hline
{$\beta_t$} & {$0.107\pm 0.045$} & {$0.093\pm 0.048$} & {$0.102\pm 0.056$} & {$0.097\pm 0.096$} & {$0.104\pm 0.059$}\\
\hline
{$\nu_t^{(1)}$} & {$0.98\pm 0.01$} & {$0.96 \pm 0.01$} & {$0.93\pm 0.01$} & {$0.92\pm 0.01$} & {$0.92 \pm 0.05$} \\
\hline
{$\nu_t^{(2)}$} & {$1.01\pm 0.04$} & {$1.0 \pm 0.3$} & {$0.95\pm 0.03$} & {$0.84\pm 0.16$} & {$0.92 \pm 0.05$} \\
\hline
\end{tabular}
\end{table*}
\end{widetext}
\end{document}